# Adaptive Attacker Strategy Development Against Moving Target Cyber Defenses

M. L. Winterrose, K. M. Carter, N. Wagner, and W. W. Streilein
MIT Lincoln Laboratory
Lexington, MA
{michael.winterrose, kevin.carter, neal.wagner, wws}@ll.mit.edu[1]

## ABSTRACT

A model of strategy formulation is used to study how an adaptive attacker learns to overcome a moving target cyber defense. The attacker-defender interaction is modeled as a game in which a defender deploys a temporal platform migration defense. Against this defense, a population of attackers develop strategies specifying the temporal ordering of resource investments that bring targeted zero-day exploits into existence. Attacker response to two defender temporal platform migration scheduling policies are examined. In the first defender scheduling policy, the defender selects the active platform in each match uniformly at random from a pool of available platforms. In the second policy the defender schedules each successive platform to maximize the diversity of the source code presented to the attacker. Adaptive attacker response strategies are modeled by finite state machine (FSM) constructs that evolve during simulated play against defender strategies via an evolutionary algorithm. It is demonstrated that the attacker learns to invest heavily in exploit creation for the platform with the least similarity to other platforms when faced with a diversity defense, while avoiding investment in exploits for this least similar platform when facing a randomization defense. Additionally, it is demonstrated that the diversity-maximizing defense is superior for shorter duration attacker-defender engagements, but performs sub-optimally in extended attacker-defender interactions.

## ABOUT THE AUTHORS

**Dr. Michael L. Winterrose** is a researcher in the Cyber Systems and Technology Group at MIT Lincoln Laboratory. He is primarily interested in developing models and techniques to aid in the understanding and shaping of adversarial dynamics observed in the cyber domain. Dr. Winterrose's research interests include advanced simulation techniques, game theory, complex systems modeling, and artificial intelligence with an emphasis on learning.

**Dr. Kevin M. Carter** is an Assistant Group Leader in the Cyber Systems and Technology Group at MIT Lincoln Laboratory. He leads efforts focused on developing models and analytics for the purposes of network security, situational awareness, anomaly detection, and decision support. His research interests include statistical signal processing, pattern recognition and machine learning applied to cyber network and system data.

**Dr. Neal Wagner** is a researcher in the Cyber Systems and Technology Group at MIT Lincoln Laboratory. His focus lies in developing and applying computational intelligence techniques for problems in the cyber domain. Specifically, he is interested in bio-inspired and heuristic algorithms for real-world scale applications of optimization, prediction, and simulation.

**Dr. William Streilein** is an Assistant Group Leader in the Cyber Systems and Technology Group at MIT Lincoln Laboratory where he manages research and development efforts focused on delivering capabilities and technologies for cyber reasoning and response. His research interests include machine learning and modeling and simulation, especially as applied to problems in cybersecurity, security metrics, and cyber moving target.

---

# Adaptive Attacker Strategy Development Against Moving Target Cyber Defenses

M. L. Winterrose, K. M. Carter, N. Wagner, and W. W. Streilein

MIT Lincoln Laboratory

Lexington, MA

{michael.winterrose, kevin.carter, neal.wagner, wws}@ll.mit.edu

INTRODUCTION

Today cyber defenders are at a systematic disadvantage in cyber conflict. Attackers often only need to exploit a single security vulnerability to succeed with an attack, and attackers can typically act at a time and place of their choosing. Furthermore, the technological monocultures that dominate information technology today place these systems at significant risk for attack. With a large number of organizations and individuals using essentially identical hardware, operating systems, and application software, significant incentives have been created for cyber attackers to discover and exploit vulnerabilities in these systems.

In this context, new techniques are under development by the cyber security research community to rebalance the playing field for cyber defenders. A major effort in recent years along these lines is an attempt by cyber defenders to diversify the most vulnerable pieces of the existing large cyber monocultures. These techniques, aiming to increase the diversity of a system's attack surface, causing increased operational costs and uncertainty for attackers have come to be grouped under the umbrella term *moving target*. Moving target techniques have been applied to diversify runtime environments, software, networks, platforms, and data in recent years (Okhravi et al., 2013; Okhravi, Hobson, Bigelow, and Streilein, 2014).

In this study, we examine a class of *migration-based* techniques that dynamically change the platform (i.e., operating system (OS)) that is active on a host in order to reduce attacker success and increase attacker resource investment requirements. These techniques work under the assumption that the attacker has limited resources and generally does not have exploits available for *all* OSes. As such, migrating between OSes with some frequency reduces the ability of an attacker to maintain persistence on a system. Additionally, it increases the uncertainty for an attacker that aims to expend resources toward exploit development.

Two recent studies have examined the optimal scheduling policy for a temporal migration moving target defense. In the first the conclusion was drawn that a uniform random scheduling policy by a defender employing a set of active spam filters performed optimally against an adaptive adversary (Colbaugh and Glass, 2012). The second set of studies (Carter, Okhravi, and Riordan, 2013) required the attacker to maintain persistence in a system for a period of time before reward accrued. An additional factor incorporated in the second set of studies was that of coupled exploits, in which a given exploit targeted at a specific OS may work against other similar OSes. It was shown that a deterministic scheduling policy that maximizes the diversity of the platforms played in each successive round was superior under the assumption of coupled exploits and the requirement of attacker persistence (Carter et al., 2013).

The goal of this work is to evaluate different scheduling policy strategies against a non-deterministic, adaptive attacker. As opposed to (Colbaugh and Glass, 2012) and (Carter et al., 2013), which posited a restrictive attacker model in which an attacker could only develop exploits for OSes presented to it by the defender, we extend our prior work (Winterrose and Carter, 2014) to model an adaptive attacker with a less restrictive attacker model. This more flexible adversary model allows an attacker to invest in the development of zero-day exploits against *any* potential defender system. Attackers observe defender actions and use these observations to learn optimal investment strategies. We will demonstrate the complex, yet intuitive, strategies that are evolved to optimize attacker success against various defender platform migration scheduling policies.

The major contributions of this work are as follows:

1. We employ a less restrictive attacker model that more accurately captures the resource investment decision problem faced by an attacker.

2. We present the first use of a novel finite state machine (FSM) construct that transitions between action states based on a heterogeneous set of system observations.

3. We show that learned attacker strategies are highly sensitive to the statistical characteristics of the defender's moving target scheduling policy.

4. We demonstrate that the degree to which a defense policy is optimal against an adaptive adversary changes as the duration of conflict varies.

**METHODS**

**Attacker/Defender Game Scenario**

Many security scenarios can be modeled as games (Tambe, 2012). Typically this involves the reduction of a given adversarial situation to its most essential elements, casting the salient features of a security conflict in stark relief. Done well, this procedure facilitates the discovery of the deeper mechanisms underlying a real-world phenomenon by eliminating the non-essential aspects of a situation.

In our scenario the attacker, characterized by a population of N strategies, plays a series of games made up of sets of consecutive matches against the defender. A simulated game between the attacker and the defender is played in which time advances in discrete steps. A single match is executed at each tick of the simulation clock. In each match, a deterministic defender activates one platform according to a pre-determined defense strategy. The attacker observes the defender's choice and must decide how to allocate resources in the next round to bring exploits into existence so as to attack the defender's system with an optimal chance of success. All exploits in this study are assumed to be *zero-day* exploits, meaning that they are unobserved by the defender when used against the corresponding operating system.

In the simulated game there exists one possible zero-day exploit for each type of platform the defender might deploy in the temporal platform migration defensive system. In each match of the game, the attacker may choose to use its resources to further develop one of these zero-day exploits.

The attacker resources (i.e., the number of rounds of attacker resource investment) required to bring a given zero-day exploit into existence is determined by sampling from a Gamma distribution at the beginning of each generation. The attacker is not informed *a priori* of the number of resources that will be required to bring a zero-day exploit against a given platform into existence. Instead the attacker discovers this only after having successfully created the exploit through the allocation of sufficient resources. Additionally the attacker must learn when to discontinue investment in the creation of a particular exploit once the required number of resources has been invested, as continued investment will be wasted. This is used to model the fact that a real-world attacker is not generally able to predict *a priori* the level of effort that will be required to develop new exploits against a given system.

In each match, the attacker uses any exploits that have been developed in the current game against the activated platform. Success for the attacker in a match occurs when it has an available exploit that works against the platform activated by the defender. Intuitively, the attacker gains a reward if the attacker is able to compromise the defender's system during the match, and earns nothing otherwise. For the purposes of this study, we impose a *persistence* requirement on the adversary, such that a reward is granted only after 3 consecutive successful matches. One may view this as the requisite length of time to stage an attack, such as the exfiltration of data over a difficult channel, with a cumulative reward being granted each match while the full attack is successful (i.e., ≥ 3 matches) (Carter et al., 2013).

In this study we allow the defender access to the pool of 5 platforms: Fedora on x86, Gentoo on x86, Debian on x86_64, FreeBSD on x86, and CentOS on x86. In the randomization defense, in each match an OS is selected from the pool of 5 for activation on the defender's system uniformly at random, with the caveat that the OS activated in the present match cannot be activated again in the immediate next match. The diversity defense consists of the deterministic activation of Fedora, Debian, and FreeBSD in succession (Carter et al., 2013). This rotation between 3 platforms maximizes the diversity of the source code presented to the attacker from match to match and reduces the likelihood that an exploit developed for one OS can persist when that OS is replaced by the next OS in the rotation.

The attacker can develop a targeted exploit for each of the defender operating systems. A *targeted exploit* works with certainty each time it is used against the platform it targets. In our model a developed exploit also works against platforms other than the exploit's target system with a probability proportional to the code similarity of the two operating systems. We term this effectiveness of an exploit against systems other than the target system its *cross platform* effectiveness.

Table 1 lists a set of *code similarity scores* for the defender's operating systems. These similarity scores were calculated using the *Measure of Software Similarity* (MOSS) tool (Schleimer, Wilkerson, and Aiken, 2003) and are based on each operating system's kernel code and standard device drivers (Carter et al., 2013). The similarity scores are given on a scale from 0 to 1, with 1 implying identical operating system code and 0 indicating completely dissimilar operating system code.

Table 1. Platform similarity scores based on operating system kernel code and standard device drivers, reproduced from (Carter et al., 2013).

|  | CentOS | Fedora | Debian | Gentoo | FreeBSD |
|---|---|---|---|---|---|
| **CentOS** | 1.0000 | 0.6645 | 0.8067 | 0.6973 | 0.0368 |
| **Fedora** | 0.6645 | 1.0000 | 0.5928 | 0.8658 | 0.0324 |
| **Debian** | 0.8067 | 0.5928 | 1.0000 | 0.6202 | 0.0385 |
| **Gentoo** | 0.6973 | 0.8658 | 0.6202 | 1.0000 | 0.0330 |
| **FreeBSD** | 0.0368 | 0.0324 | 0.0385 | 0.0330 | 1.0000 |

We note that FreeBSD is an outlier in the set with a markedly low similarity score compared with the remainder of the set. This is explained by the fact that FreeBSD is based on Unix while the other 4 operating systems are Linux-based. In the sections that follow the outlier status of the FreeBSD platform will be shown to have a significant impact on attacker strategy development.

In this study the cross-platform effectiveness is determined on a match-by-match basis. On each match the attacker plays all available exploits against the defender activated platform If the attacker has developed the exploit *targeted* at the activated defender's system then the attacker succeeds with the attack with certainty. On the other hand, if the targeted exploit has not been created by the attacker for the activated platform, any other created exploits succeed against the defender's system with a probability equal to the similarity score between the exploit's targeted system and the system activated by the defender.

**Finite State Machine Strategy Encoding**

We represent attacker strategies as binary chromosomes encoding a FSM construct. An FSM is an abstract machine that takes discrete inputs from an environment and specifies a discrete output in response. An agent modeled by an FSM will occupy only one state at any point in time. Such an agent transitions between states based on observations of its environment.

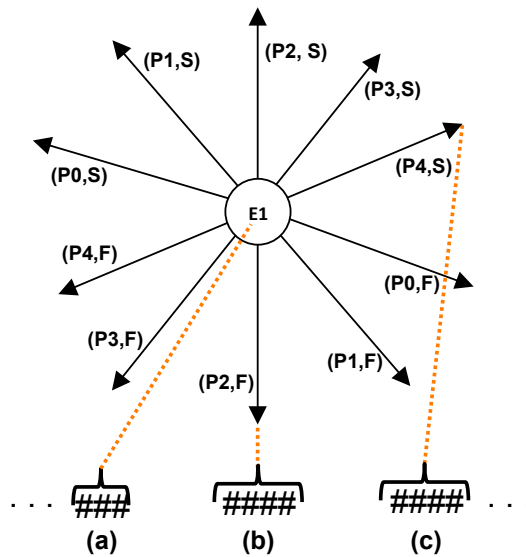

**Figure 1.** Hypothetical single automata state and corresponding outgoing transition set of an attacker's FSM (upper). *E, P, S,* and *F* represent *Exploit investment*, *Platform observation*, *Successful attack*, and *Failed attack* by the attacker, respectively. The finite automata state maps into a binary chromosome in which bits are represented by # ∈ {0, 1}. Portion (a) of the chromosome encodes the attacker investment in zero-day exploit creation when the attacker occupies this hypothetical state using 3-bits. Segments (b) and (c) of the chromosome encode two possible transitions executed in response to observation of the defender's activations and the attacker's success in the current round. The transitions are encoded using 4-bits each

Each strategy in our 30-strategy population is represented by one 16-state, 160-transition FSM. Each state encodes up to 8 possible actions, leading to a 692-bit chromosome encoded in a manner similar to (Miller, 1996; Winterrose, 2014). Figure 1 depicts a single state in our machine and its outgoing transitions.

The 30 machines are initialized randomly before the simulation begins. During the simulation the actions encoded in each state of each machine and the transitions between states evolve according to the genetic algorithm presented in the next section.

We use 16-state FSMs for historical reasons (Miller, 1996), but find through ancillary studies that the actual strategies evolved by attackers generally fit easily within our 16-state constraint. An extended study with a widely varying number of machine states would shed useful light on the consequences of *bounded rationality* on the nature of strategies evolved in the cyber domain.

Our FSMs transition between states based on both the type of platform activated by the defender in the previous round and on the success the attacker had with its exploit attacks in the previous round. To the best of our knowledge, this *dual-observation* transition model is unique to this study. Previous studies using a simpler, but related, FSM construct to play the Prisoner's Dilemma game-theoretic scenario (Miller, 1996) transitioned between machine states based on a single observation of opponent action in each round of play.

**Evolutionary Algorithm**

The adaptive attacker in our study evolves strategies against the defender using a genetic algorithm (GA) (Holland, 1975). Originally conceived as a stylized model of biological evolution, the GA has proven to be a robust method that can efficiently search solution spaces that are nonlinear and/or discontinuous. In our implementation we randomly initialize 30 strategies at the outset of a simulation run. In each generation each *agent* (i.e., *strategy*) plays a *game* consisting of M matches against the defender.

In a *match*, if the defender's active platform is vulnerable to an exploit that has been successfully developed, the attacker accrues a reward, governed by some underlying function that is hidden from the algorithm. This may include immediate reward, or for example, require some level of consecutive success before a reward is granted (e.g. *persistence*). See (Carter et al., 2013) for example scenarios and associated reward functions.

Once the reward is computed, the match is concluded. A new match begins with the attacker choosing an exploit to develop with its allocated resources (one resource is available for investment by the attacker in each match). Concurrently, the defender selects a platform to make active in the system, against which the attacker moves with any available exploits. This continues for M matches, at which point the *game* ends between the chosen attacker and the defender, and a new attacker strategy from the population is rotated in to play against the defender. Once all attackers have played their M matches in the generation (g) against the defender, each strategy, *i*, is assigned a fitness score, *F*, based on its success against the defender,

$$F_{i,g} = \sum_{j=1}^{M_g} \Phi_{i,j}, \qquad (1)$$

with $\Phi_{i,j}$ set equal to +1 if the system is compromised, and set equal to 0 otherwise. Attacker strategies are ranked based on their fitness scores. A new population of attacker strategies is generated for play against the defender using the following steps:

1. A fraction of the top ranked attacker strategies are copied directly into the new population. This procedure is known as *elitism* and is commonly used in GA applications to avoid the loss of the best strategies from previous strategy populations (Mitchell, 1996).

2. Two attacker strategies are chosen from the current population using fitness proportionate selection in which higher ranking strategies are more likely to be selected.

3. The two selected (parent) strategies undergo the *crossover* genetic operation (analogous to biological sexual reproduction) to generate two offspring strategies. In this operation, a single crossover point c ∈ {1,2, . . . n} on each of two parent chromosomes is selected uniformly at random. The first offspring combines the first c bits from the first parent with all bits after the c+1 chromosome position of the second parent to form a new chromosome. The second offspring takes all bits after the c+1 chromosome position from parent 1 and combines it with the first c bits of parent two's chromosome to form a new strategy.

4. The two offspring strategies are then subject to the *mutation* genetic operation (analogous to asexual reproduction). In this operation bits in the chromosome are randomly altered. The mutation operation is commonly used in GA applications to increase population diversity and avoid local extrema in the search space (Michalewicz, 1996).

5. The two generated offspring strategies are then added to the new population.

6. The above steps are repeated until the new population has a sufficient number of strategies (specified by a population number parameter).

Attacker strategies are evolved over a set of generations where each generation includes the attacker-defender simulated games and the above steps to generate new populations of strategies.

**EXPERIMENTS**

**Simulation Initialization**

The defender's dynamic platform scheduling policy is assigned at the beginning of the game and is not altered as the game progresses. The attacker's strategy is represented by a population of randomly initialized strategies encoded as binary chromosomes representing FSMs. Each iteration of the simulation is allowed to run for 100 generations of genetic algorithm evolution, with the attacker strategy being evolved in each of these generations. A single generation consists of each of the N=30 attackers playing M matches against the defender. The 100-generation run is iterated 100 times and the results aggregated and averaged to account for the stochasticity in the model.

The number of attacker resources required to bring a given zero-day exploit into existence is determined at the beginning of each generation by independent draws from a Gamma distribution for each of the 5 possible zero-day exploits available for development by the attacker, similar to the procedure we used in (Winterrose and Carter, 2014). The Gamma distribution is parameterized by a mean (μ) and variance ($\sigma^2$) parameter. We use μ =25 and $\sigma^2$=10 throughout this study.

In the analysis that follows we typically extract the fittest learned attacker strategy in each generation of each simulation run and aggregate these together to produce the results discussed. This procedure is consistent with the focus in this paper on the nature of the optimal attacker strategies developed against the defender's moving target defense. We refer to this set of fittest strategies extracted from each simulation run as the *fittest attackers* or *fittest strategies* hereafter.

Simulations were created and executed in the NetLogo modeling environment (Wilensky, 1999). Data aggregation across simulation runs and the calculation of statistical measures was carried out using MATLAB release 2013b (Matlab, 2013). To visualize the evolved FSMs we have used the Gephi network visualization and analysis software package (Bastian, Heymann, and Jacomy, 2009).

**Attacker's Response to Diversity and Randomization Defense**

Figure 2 shows the match-level response at the 100th generation of genetic algorithm evolution for the fittest attackers averaged over 100 simulation runs. The attacker performs better against the randomization defense throughout the 100-match game, but the difference in performance narrows as the match number increases. We recall that the attacker must compromise the defender's system for 3 consecutive matches using its developed exploits before beginning to accrue a reward for system compromise. Figure 2 shows that this begins to occur at an earlier point in match play when the attacker faces the randomization defense. Specifically, the attacker's fitness begins to rise around match number 40 when the attacker faces the randomization defense, roughly the point at which an efficient attacker might begin to have access to 2 exploits given this studies' exploit creation cost parameterization. With 2 exploits created the attacker can utilize the cross-platform effectiveness of the created exploits to achieve the persistence required for accruing attacker reward.

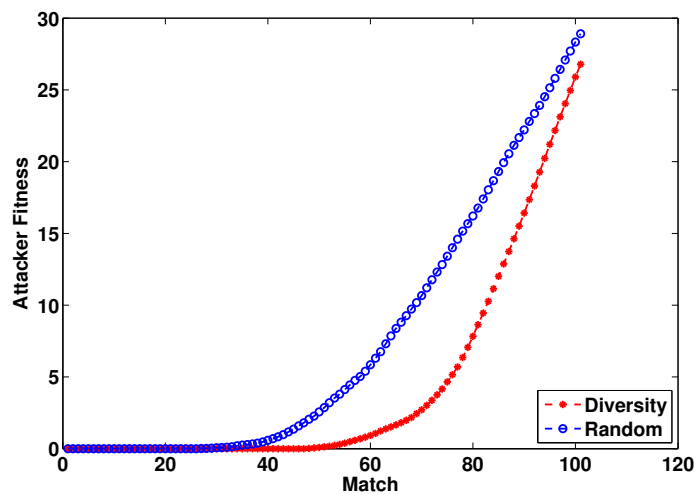

**Figure 2. Fittest attacker game success in the 100th generation of genetic algorithm evolution averaged over 100 simulation runs. The attacker is most successful against the defender deploying the randomization dynamic platform scheduling policy, though the difference in response narrows in later matches. See text for discussion.**

Against the diversity defense, on the other hand, the attacker does not begin accumulating reward until just before match 60. Between the 60th and 75th match the attacker's reward (i.e., fitness) climbs slowly, then accelerates sharply after approximately match 75. This can be understood by recalling that once the attacker has had the opportunity to develop 3 targeted exploits it is able to completely counter the diversity defense. This causes the fitness of the attacker facing the diversity defense to quickly approach the fitness of the attacker facing the randomization defense.

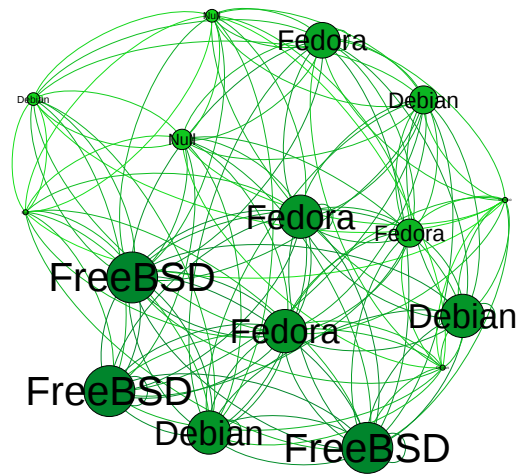

**Figure 3. Structural representation of an exemplar attacker strategy developed to counter the diversity defense. The strategies are encoded in a FSM. Nodes are labeled by the investment to be made by the attacker in the various machine states. Edges represent transitions between states based on observations of the defender's actions and successful game play. See text for further discussion.**

Figure 3 shows the structural properties of an exemplar FSM encoding an attacker's evolved strategy when facing the diversity defense. In the figure the node and label sizes are proportional to the number of transitions into a given state. The importance of the Fedora, Debian, and FreeBSD exploit development in the learned attacker strategy are clear in this FSM representation. In particular, FreeBSD is the most prevalent investment state in the structure, a fact we discuss in the next section.

**Patterns of Attacker Investment in Zero-Day Exploit Creation**

An important consideration when deciding upon a deployment strategy for a dynamic platform moving target defense is how the attacker is likely to alter its strategy based on the defender's choices. For this experiment, we were interested in understanding how the statistical characteristics of the defender's scheduling policy affects attacker exploit creation investment choices.

The basic choice the attacker faces is the manner in which to invest its resource in each round to compromise the attacker's system with maximum effectiveness. The key considerations for the attacker in achieving this goal is the persistence requirement (i.e., 3 consecutive successful attacks before attacker reward accrues) and the cross-platform effectiveness of each zero-day exploit. The need to weigh these factors together with the observations of the defender's dynamic platform scheduling policy make the investment choice a complex one for the attacker.

Figure 4 shows the generational investment patterns learned by the fittest attackers aggregated across the 100 simulation runs. It is clear that the statistical character of the defender's scheduling policy strongly affects the exploit investment pattern of the attacker. The largest effects are observed in the preference or disdain the attacker shows for developing the FreeBSD zero-day exploit. Figure 4a shows that when facing the randomizing defender the attacker prefers to minimize investment in the FreeBSD exploit and focus investment on the creation of exploits for the Linux-based platforms. This behavior contrasts sharply with the attacker's response to the diversity-maximizing defender (Fig. 4b). Here the attacker shows a strong preference for developing the FreeBSD exploit.

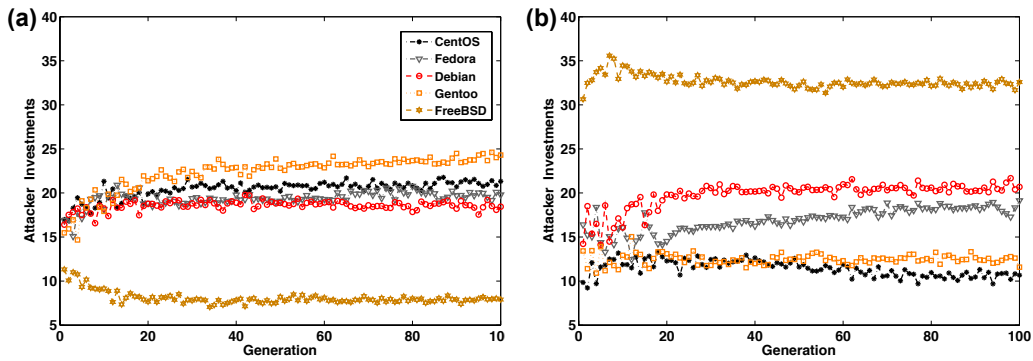

**Figure 4. Generational progression of fittest attacker exploit creation investments averaged over 100 simulation runs. Investment patterns of attackers facing a randomized defender scheduling policy (a) differ markedly from the investment patterns developed by the evolving attackers facing a diversity defense (b).**

We note that in Fig. 4 the attacker has discovered these investment patterns already in the initial generation. This early discovery of the fittest attacker strategy is essentially a matter of luck. The process of learning these investment patterns is more evident when examining the evolution of investment patterns within the entire population of N=30 attackers, as shown in Fig. 5. Here we see the population mean investment in each of the available exploits distributed approximately uniformly in the initial generation, then diverging strongly in just a few generations as the population converges on the strategies of avoiding FreeBSD exploit investment when facing the randomized defense (Fig. 5a), and investing heavily in FreeBSD exploit creation when facing the diversity defense in Fig. 5b.

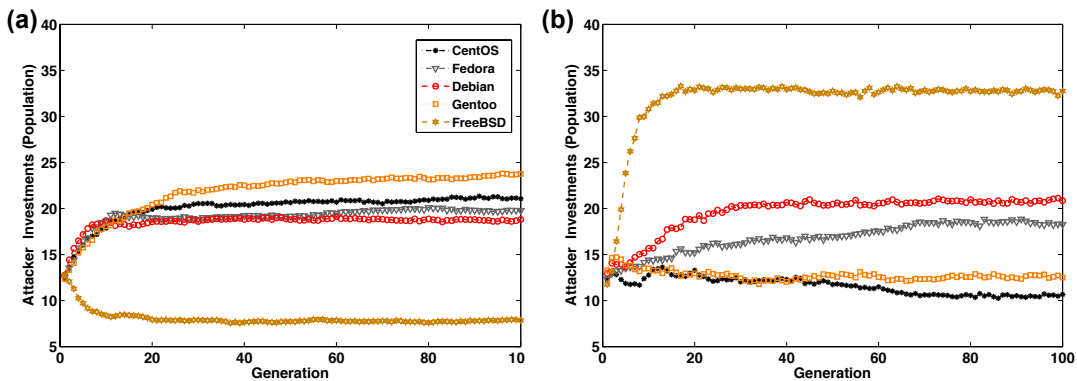

**Figure 5. Generational progression of mean attacker exploit creation investments for the entire attacker population of N=30 strategies. Displayed results are aggregated and averaged over 100 simulation runs. As in the fittest attacker case, investment patterns of attackers facing a randomized defender scheduling policy (a) are quite different from the investment patterns developed by the evolving attackers facing a diversity defense (b).**

These trends can be understood by taking account of the following observations. When facing the diversity defense the attacker can predict with certainty that it will face a defender activating the FreeBSD platform reliably every 3 matches. Given the dissimilarity of FreeBSD with the other 4 platforms, this makes it improbable that the attacker will achieve the requirement of 3-match persistence across the FreeBSD activation if the FreeBSD exploit has not been created. When the attacker is facing the randomization defense, in contrast, there exists a reasonable probability that the attacker will achieve the persistence requirement and accrue reward without facing activation of the FreeBSD platform by the defender. The predictability of needing to overcome a FreeBSD activation in the first case (i.e., diversity defense), and the uncertainty of facing a FreeBSD activation in the second case (i.e., randomization defense) rationalizes the investment patterns in Fig. 4 and 5.

**Engagement Duration Effects**

Another important consideration when deciding upon a deployment strategy for a dynamic platform moving target defense is the duration of the interaction. For this experiment, we were interested in understanding how attacker fitness in the face of diversity and random strategies was affected by different interaction (i.e., game) durations. Figure 6 demonstrates that the duration of attacker/defender interaction greatly affects the efficacy of the defensive capability, which is reflected inversely in the fitness level of the attacker: better attacker fitness implies worse defensive capability. In Fig. 6a we see that the attacker achieves a high level of fitness for a 75-Match game when a random strategy is utilized. By contrast, when the defender utilizes the diversity strategy, the attacker never achieves a similar level of fitness, though the overall level of fitness does increase with generation. As the duration of the games increases, through 100-Match and 125-Match games, Fig. 6 shows that the value of the diversity strategy diminishes for longer duration interactions. Specifically, when 100-Match games are played, the attacker fitness level is roughly equivalent for random and diversity strategy at the start, with the diversity strategy initially performing sub-optimally and then improving. When the 125-match games are played, the diversity strategy is always sub-optimal to the random strategy in providing effective defense by allowing the attacker fitness to reach a higher level. It is worth noting that the absolute fitness level achieved by the attacker increases overall as the interaction duration increases regardless of the defensive strategy employed. This is due to the fact that the attacker is provided with more time to develop an exploit in all cases regardless of the defensive strategy in use and thus is able to improve fitness level.

To understand the benefit provided at shorter durations by the diversity strategy it is instructive to consider that although the attacker is able to focus his resources on a smaller set of target OSes, the shorter duration makes it difficult to achieve all exploits needed to compromise the system with the required persistence. As the duration increases, the attacker is more likely to develop targeted exploits for the entire set of the OSes in the diversity strategy before the interaction ends and thus is able to achieve persistent compromise. When the game duration reaches 125 matches, the random strategy provides more defensive benefit due to the attacker's increased difficulty in predicting future OSes relative to the diversity strategy.

As a result of these simulated experiments, it is recommended that when interaction with an attacker can be kept to a short duration, a diversity strategy is preferred. The specific length of the duration that is optimal depends upon the expected time it would take the attacker to develop the exploits.

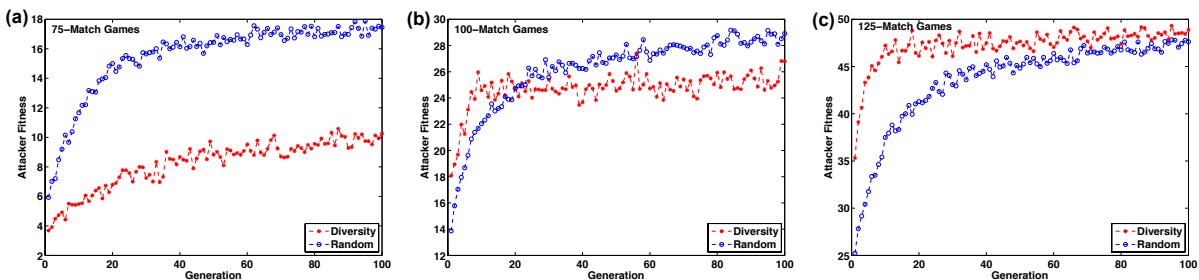

Figure 6. Game success as a function of generation number for the set of fittest attackers facing the defender in games of varying length, as indicated in the figures.

**CONCLUSION**

We have developed a model of adaptive attacker strategy evolution and used it to investigate the strategies an attacker develops to overcome two temporal platform migration moving target defense strategies. The attacker-defender interaction has been modeled as a game in which a non-adaptive defender deploys a randomization or a diversity moving target defense. Against these dynamic platform scheduling policies a population of attackers develop strategies specifying the temporal ordering of resource investments that bring zero-day exploits into existence to compromise the defender's system.

The results of this study have strong implications for real-world defenders. First, defenders deploying dynamic platform defenses and anticipating attacks over *difficult channels* (i.e., requiring persistence to succeed), should be particularly vigilant regarding the systems in their rotation-set with outlier status in attributes relevant to an attacker's success. It is these outlier systems that advanced attackers will devote the largest proportion of resources to compromising. Furthermore, our results suggest that diversity-maximizing defenses are most effective for short duration attacker/defender encounters. The crucial parameter in this regard is the time required for an attacker to bring exploits into existence versus the duration of the attacker's encounter with the defender's system.

Future directions of interest for investigation include the incorporation of noise into the attacker's observation model in order to bring the game scenario nearer to conditions likely to prevail for real-world attackers and the incorporation of an adaptive defender into our cyber game scenario.